# A knowledge engineering method for new product development

Nicolas Perry*, Samar Ammar-Khodja**
1 LGM²B, Univ. Bordeaux 1, Bordeaux, France
2 Glaizer Group, Malakoff, France

**Abstract**

Engineering activities involve large groups of people from different domains and disciplines. They often generate important information flows that are difficult to manage. To face these difficulties, a knowledge engineering process is necessary to structure the information and its use. This paper presents a deployment of a knowledge capitalization process based on the enrichment of MOKA methodology to support the integration of Process Planning knowledge in a CAD System. Our goal is to help different actors to work collaboratively by proposing one referential view of the domain, the context and the objectives assuming that it will help them in better decision-making.

## *1. Introduction*

In recent years, engineering systems have moved from being information-intensive towards knowledge-intensive systems [1] [3]. The information is thus constantly refined by clarifications, discussions and evaluations, until an optimised or compromised solution is agreed. In the framework of Product design and manufacturing, Weber [5] argue that today's PDM and PLM systems provide infrastructures to store and move data, but not retain knowledge about the content and the interrelationships of the data they handle. But, decisions are based upon the designers' intellectual assets and vary from one expert to another.

It becomes crucial to develop a method for take the benefit of this intellectual capital. The Knowledge-Based Systems (KBS) are one solution. Their development relies on the transformation of human informal knowledge into formal knowledge with some support from knowledge engineering techniques [6].

The purpose of this paper is to introduce an engineering process to structure the transfer of expertise from experts' minds to an automated system in the manufacturing domain. Assuming that capitalizing knowledge consists of working on the content and form of the knowledge, the proposed process is structured in two major phases: the capture phase and the formalisation phase. Focusing on the first phase, the main objective of this work is to define a capitalization process to support knowledge capture and representation for the specification of a knowledge-based engineering (KBE) system, which is a specific type of knowledge-based systems.

This study is based on the USIQUICK project which. First, we will start by describing its global context and objectives. Then, the problematic will be presented before developing the capitalization process proposed.

## 2. Knowledge-based methods and tools

A knowledge-based system can be defined as a computerised system that uses knowledge about some domain in order to deliver a solution concerning a problem [7]. The first generation of knowledge-based systems was expert systems using a set of facts and rules [8]. This kind of systems is composed of essentially two components: a knowledge base (KB) and an inference engine. It applies specific domain or domain-specific knowledge to problem-specific data to generate problem-specific conclusions [9]. The next KBS generation was the case-based systems. These systems use previous solutions to problems as a guide to solving new problems. Knowledge-based systems are widely acknowledged to be the key for enhancing productivity in industry, but the major bottleneck of their construction is knowledge acquisition, i.e. the process of capturing expertise before implementation in a system [13]. Some methodologies assist the developers in defining and modelling the problem in question, such as Structured Analysis and Generation of Expert Systems (STAGES) and Knowledge Acquisition Documentation System (KADS) (an acronym that has been redefined many times, e.g. Knowledge Acquisition Documentation System and Knowledge-based system Analysis and Design Support). Moreover, these approaches get enriched in order to take into account the project management, organisational analysis, knowledge acquisition, conceptual modelling, user interaction, system integration and design [14] [15]. Consequently, knowledge modelling in engineering must be based on a rich and structured representation of this knowledge, and an adequate way of user interaction for modelling and using this knowledge [16]. Due to the complexity of engineering knowledge, knowledge modelling in engineering is a complex task.

KBE has been defined as being an engineering methodology in which knowledge about the product, e.g. the techniques used to design, analyse, and manufacture a product, is stored in a special product model. The product model represents the engineering intent behind the geometric design. The KBE product model can also use information outside its product model environment such as databases and external company programs. KBE has been defined as "a computer system that stores and processes knowledge related to and based upon a constructed and computerised product model" [7]. The encoding of design knowledge from domain experts into computer codes that can generate complex geometric data, has demonstrated significant savings in manpower and time resources for routine design problems [17], and has also provided a high degree of design integration and automation in well-defined and complex design tasks. The MOKA methodology has been proposed to address methodological issues during KBE systems development for our case study.

The modelling approach in KBE has to structure the engineering knowledge. In terms of developing KBE applications, this structuring process involves the configuration of the objects that model the engineering design environment and the rules that control the behaviour of the objects [1]. Current KBE systems are based upon a combination of the production rules and the object-oriented knowledge representation. Both elements together offer an automated way to introduce design requirements, model design constraints and provide a product description.

## 3. Knowledge base project: USIQUICK

Engineering knowledge tends to be very complex, diverse, and interrelated in many ways. Consequently, knowledge modelling in engineering must be based on a rich and structured representation of this knowledge, and an adequate way of user interaction for modelling and using this knowledge [16]. Still, due to the complexity of engineering knowledge, knowledge modelling in engineering is a complex task. Many relations and interdependencies have to be taken into account in order to come up with a model that is as precise, generic, consistent and concise as possible [1]. So, each new piece of knowledge, which should be inserted into an existing knowledge model, has to be related in many ways to the already contained knowledge. Thus, during modelling, a maximum of information about the already existing model has to be available and easily accessible by the knowledge engineer.

The other main knowledge-related issue in engineering is the application of knowledge-based technologies, i.e. the automatic computer-based processing of knowledge in KBE systems.

The following two sections define the concept of KBE, the most well-known methodologies and most-widely-used modelling techniques to support such technology.

### a. Context

The works presented are part of the output from an industrial project (USIQUICK [25]). The project aimed at developing a knowledge-based engineering system to help experts during the process planning for mechanical parts. The project involves eight partners. An aircraft manufacturer is the final user and the initial expert. He specified the expected results and its manufacturing expertise on complex part design and on process planning. A CAD/CAM developer supported the industrialization in its software solution. Five laboratories ensured the scientific coherence, enriched and solved the strategic keystones of the project. A French-government institute helped to switch these project results in other area of mechanical manufacturing. The partners started working together in a same setting domain with different cultures, contexts, goals and backgrounds. These differences led to different viewpoints, assumptions and needs. Furthermore, they used different jargons and terminologies sometimes diverging or overlapping, generating and becoming unclear.

In order to optimise the information flow from design to production, a three-step method is proposed [22]:

- Transformation phase: an analysis of the part to compute a maximum of information registered at an appropriate level of feature. In this phase computer assesses the machinability of faces.
- Preparation phase: the synthesis templates of the previous phase are presented to the user. Then with appropriate tools, the process plan skeleton can be built and constrained.
- Automation phase: the unconstrained choices are automatically optimized and a complete documentation is proposed by the system.

These phases would become the three major modules of the engineering tool based on the formalisation and the integration of expert knowledge.

We play a role in the project in order to propose solutions to allow to effectively cooperate on the same objective despite the mentioned differences, and to reduce the communication gap between the domain expert and the developer. Contextualised and structured information was shared, in the form of knowledge, to help all the actors to have a same understanding of the domain, the context and the goals.

However, to develop a KBE system, we need first to acquire, represent, reason and then communicate the intent of the design process. The problem is first understood at a conceptual level, and then decomposed into understandable working objects, developed further through an iterative process until a satisfactory outcome is reached. Then, product and process development are defined as a logical sequence of stages or activities, which may be documented, disseminated and understood by all the actors [18].

One of the project's challenges is to translate knowledge that has been expressed in the form of legacy specifications for the development of the system into a computerised form so that the computer can use it. The difficulty is thus to select the right methods and tools for supporting and structuring such a transfer. One solution could be to structure the knowledge within a knowledge base (KB) (figure 1-a). The building of this KB implies the deployment of a capitalization process to help and guide the knowledge treatments. Capitalizing knowledge consists in processing and treating knowledge to prepare it for management activities. This capitalization will enable knowledge to be shared through a specific form making it understandable by each actor of the project.

The next section details the definition of such a process and highlights its major steps. This definition represents an introduction to the process we are proposing (following section). An overview describing MOKA methodology principals and ontology will be presented.

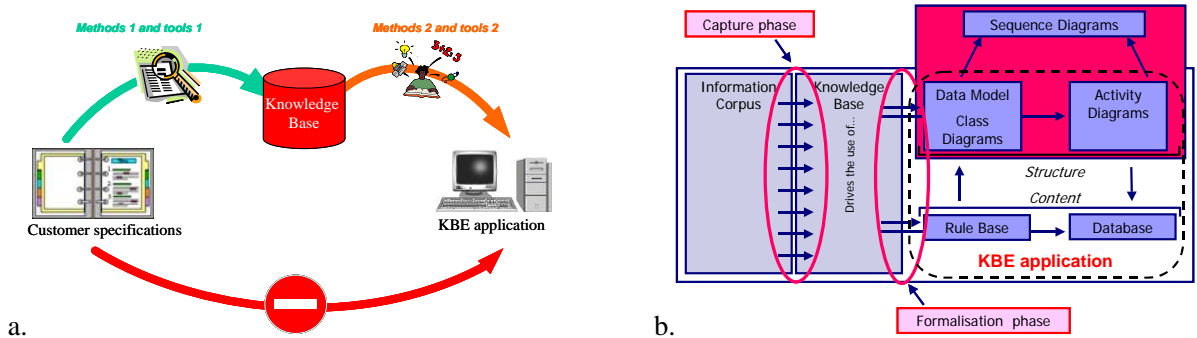

**Figure 1 : knowledge transfer possibility (a) and the knowledge capitalisation phases (b)**

### b. Knowledge capitalization

Knowledge capitalization is the process of capturing and formalising expertise before its implementation in a system. This process can be refined into four major stages:

- Knowledge elicitation, also known as acquisition, the process of obtaining knowledge from an expert;
- Knowledge analysis, the process of making sense of the information collected in the first step;
- Knowledge structuring, the process of expressing the analysed knowledge in an understandable and usable form, for enhancing communication between the expert and the knowledge engineer, and for validation purposes;
- Knowledge representation, the process of rearranging and expressing knowledge in a format that facilitates its encoding and thus its handling by a computer.

The aim of knowledge capitalization is to develop methods and tools that make the task of capturing and validating experts' knowledge as efficiently and effectively as possible. Experts tend to be important and busy people; hence it is vital that the methods used minimise the time each expert spends off the job taking part in knowledge acquisition sessions [7].

To reach the multi-experts collaboration, the knowledge sharing and reuse within the USIQUICK context, we propose to capitalize the knowledge in two major phases: a capture phase and a formalisation phase.

The capture phase gathers the elicitation, the analysis and the structuring stages, while the formalisation phase is the representation stage. In the following sections only the capture phase will be detailed.

## *4.   Capitalization process proposal*

According to the KBE systems development principle, knowledge must be identified, acquired, analysed, structured and formalised in such a way that it could be accessible and reusable by each one. However, this principle does not allow any distinction between the activities handling the knowledge content (this means knowledge itself) and those handling its form.

What we are proposing in this paper is not completely different from or contradictory to the KBE development principle. Our aim is to structure all these activities according to the knowledge aspect addressed at each stage of the capitalization process. This structuring consists in separating the activities that handle the knowledge itself from those handling its form. This distinction tends to help knowledge engineers during capitalization activities deployment.

This structuring can also be considered as working on the state of the knowledge. Working on the knowledge content consists in transforming its state from a raw state (independently of being explicit or tacit) to a structured one. Working on the form, deals with the representation of the knowledge in order to go from a structured state to a formalised state, and onwards toward an automated one.

The transition between the two phases is based on the design of a knowledge base. This base constitutes a knowledge repository that can be accessible and which will be the knowledge reference for all the partners involved (figure 1-b).

## *5.   Knowledge Capture Phase*

Knowledge capture is the process that tries to transform the human experts' knowledge into a formulated knowledge that can be used directly by an expert system or by a computer system.

As defined in the previous section, this process can be broken down into three major steps: the elicitation step, the analysis step and the structuring step.

### a.  Elicitation step

The terms "knowledge elicitation" mean "how to obtain (or collect or acquire) knowledge from an expert". Diaper [23] has extended this definition to include elicitation from other sources, such as documents, existing computer systems and the physical or the social environment.

Many elicitation techniques exist depending on the type of the knowledge source. The most common way to elicit knowledge from an expert is interviews. These interviews can be structured or unstructured depending on their context and on the knowledge engineer's strategy. On the other hand, eliciting knowledge from documents can be done by data mining techniques resulting from artificial intelligence.

Within the USIQUCK project, the elicitation had to be done from documents that represent legacy specifications for the development of the final system.

Among the existing methodologies for KBS and KBE development, the only one that can meet our needs is MOKA. This is because it offers the possibilities of eliciting knowledge from documents within engineering domains through its ontology. Ontology is a set of different interrelated concepts that describe a given domain [24]. However, this does not mean that MOKA does not allow the eliciting of knowledge from experts by using the proposed ontology.

To do so, we chose to deploy the proposed ontology within MOKA in order to identify the concepts that should be acquired from the specifications we obtained. However, before explaining this deployment, we will present MOKA.

# 1. MOKA methodology

MOKA, for Methodology and software tools Oriented to Knowledge Engineering Applications, describes in terms of rules, processes, modelling techniques and definitions, the necessary stages for the specification of KBE systems. MOKA provides a framework both for capturing and for representing knowledge. This framework works at two levels: informal level and formal level. The first one is relatively simple and oriented to represent and formalize knowledge in language that can be understood by experts without being a specialist in formalization languages. The advantage of this level is that it makes the validation of the acquired knowledge possible. This level also facilitates the communication between the expert, the knowledge engineer and the software developer.

The second level is more formal and aims to represent and store knowledge in an encoding form in order to plug it into computers.

The MOKA spirit is not different from the approaches proposed within the other knowledge management methodologies, the difference lies in the deployment strategy.

The other point that differentiates it from the other methods is the concepts it proposes to analyse the application domain. MOKA proposes five generic knowledge object types and relations among them to describe the domain. These objects as well as their use constraints are also defined. These object types are:

- Illustrations representing comments, past experiences, specific cases and complex explanations;
- Constraints describing the product's or its component's limitations;
- Activities to describe problems resolution stages;
- Rules to describe knowledge that directs the choices in the activities;
- Entities to represent knowledge elements that describe the product, its components, its assemblies, parts and features. An entity can be structural or functional.

Starting from this ontology, our first step was the identification of the knowledge objects. The identification step is a preliminary domain investigation and analysis that aims to recognise the knowledge elements or objects that must be acquired. The specifications we obtained consisted of texts, tables, and images in MS Word format. The domain library, which approximates domain ontology, consists of technical sentences condensed from legacy specifications.

The use of the MOKA ontology enabled us to identify a great number of knowledge objects. However, there is some knowledge related to, for example, resources and functions that have been missed.

The insufficiency of the ontology in this case study is due to the fact that in our context the final product is a process planning which is a process. The object's types do not become reusable as proposed. For example, if we consider the structural entity, it describes a physical component of the product but within our context the product is not a set of physical components but a set of activities that consist in geometry recognition, manufacturing mode identification, manufacturing operations definition and organization, etc. They represent domain activities. This implies that we have two types of activities, those related to the domain and those related to the reasoning that allows definition of the process planning. The reasoning activities represent the design process and each one covers one or several domain activities. This insufficiency led to a need for ontology enrichment.

**2. MOKA ontology enrichment**

Facing this insufficiency, we propose to define the concept of resource to encapsulate the knowledge of the different tools and machines used by manufacturing processes (or operations) to realize geometries. Hence, this object should be considered at the same level as the entity and the activity. It should also be related to both of them.

We also propose to define a concept of function to identify what is the objective of the reasoning activities. During the design of the system, some reasoning activities that have to be encoded aim to list results or to check if some parameters values are correct or not. This kind of activity should be attached to the concept of function to allow the differentiation of the activities related to a problem solving from those related to the presentation of the solution. It will be linked to the activity. The concept of entity in our context will represent the manufacturing features to be realized. We also distinguished the representation constraints from the product constraints and also the expert rules from the domain rules.

The representation constraints describe the constraints related to the presentation of the knowledge to the end user, and the product constraints enable the definition of all the constraints related to the product and its design.

The domain rules cover generic rules defined in the domain and the expert rules describe rules, applied by a specific expert that can vary from one expert to another.

According to these new object types, we propose ICARREF ontology to cover the manufacturing domain, in this case study, and for capturing knowledge about a product that is a process considering that these object types are generic. Figure 2 illustrates all object types and their interrelations. This figure also shows the ICARREF forms to fill in, and the ways to navigate within the knowledge base.

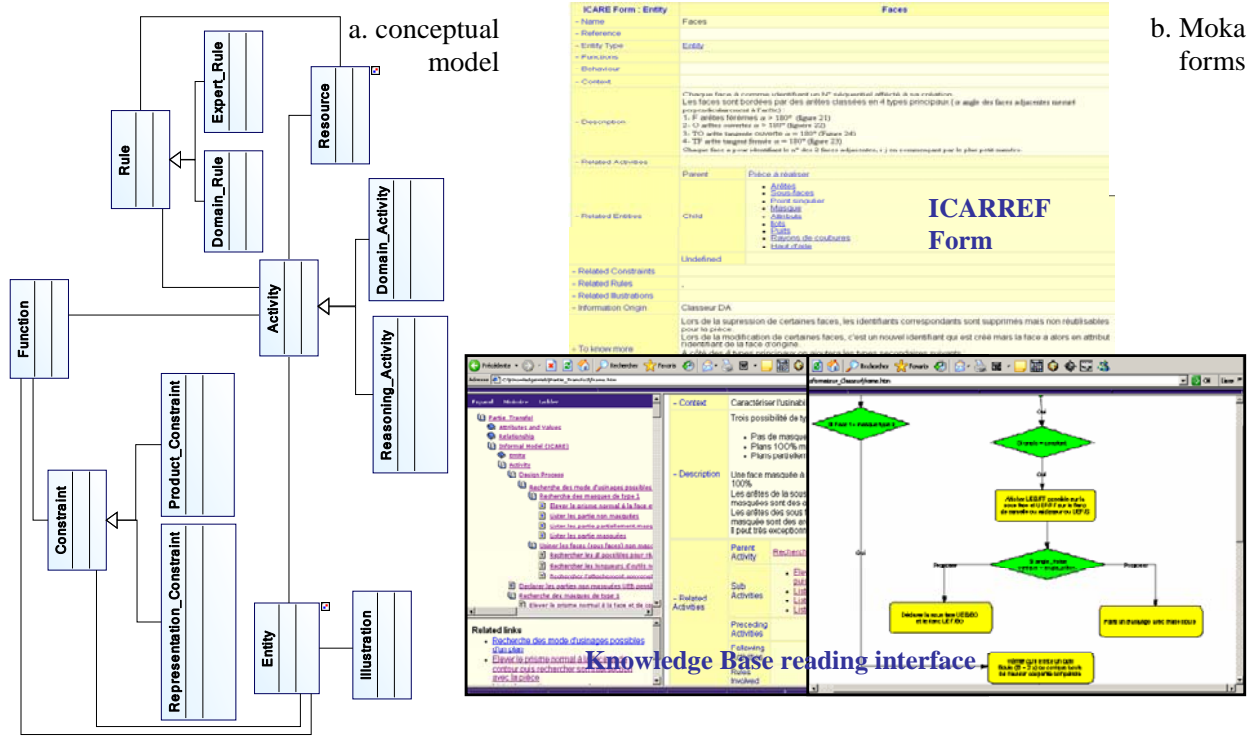

**Figure 2:** ICARREF conceptual model and working interface

At this stage the knowledge to be kept has been identified and the elicitation can be done completely by an extraction strategy. The extraction consists of recognizing a subset of knowledge objects and their relationships, and then associating them with applicable fragments of the specifications (figure 3-a). The eventual output of extraction can be in plain text, in XML, or in Excel form, depending on the application of the supported software. In this example the output is in plain text.

Once the knowledge is extracted it must be analysed. This analysis has two objectives: its structuring and its evaluation.

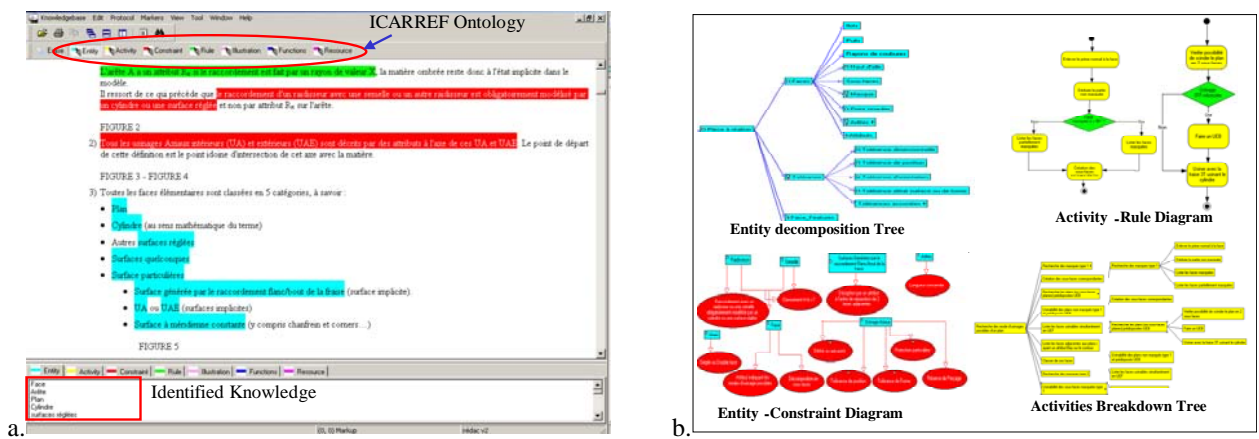

**Figure 3:** Knowledge acquisition (a) and structuring (b) step

### b. Knowledge analysis

The analysis step is the most difficult step in the Knowledge-capture process because the belief is that a "magical one-to-one correspondence" between the expert's verbal comment and the real items of knowledge is misleading [15]. Data and information obtained from manuals, textbooks, experts, and even users need to be converted into knowledge before they can be used.

The intermediate step of knowledge analysis is important, because its result will enable the building of a first knowledge model of the domain and the reasoning. It consists, first, of identifying the interrelated knowledge components, and after, on defining the right relation for each linked components.

Different relation types can be defined, for example: has constraint to link entities to constraints, has function between functions and activities, etc.

Once the interrelations and the relations have been defined, the knowledge should be structured.

### c. Knowledge structuring

The structuring step will be achieved using trees and diagrams according to the MOKA approach. Knowledge objects having the same type are linked using trees with "Is a" and/or "Is composed of" relation types. Knowledge objects having different types are linked using diagrams (figure 3-b). For diagram building, the relations are defined according to the objects they link. It can be "Has a rule", "Has a constraint", "Has a function", etc.

At this stage, the three steps of the capture phase have been done and a first representation of the knowledge is built. This representation will enable the evaluation of the knowledge.

The evaluation consists in analyzing the knowledge according to two criteria: completeness and feasibility.

The completeness indicates if, as transmitted by the expert in the specifications, this knowledge is enough to define the process planning for specified geometries. It also allows identification if, for a specified utilisation of the application, the context for each knowledge object is well described. This criterion highlights the additional knowledge to capture or to explain further if this has already been done.

Each one knows that there is a gap between the "real world" and the "computer world". The analysis of the feasibility to point out the knowledge that cannot be coded as specified by the expert and that requires the development of additional algorithms to make its automation possible.

## *6.    Knowledge completeness*

Figure 4 illustrates the automatic semantic enrichment of surfaces that will be machined. The type of these surfaces (colour identification in 4.a) depends on the rules and constraints linked to the tools access, machining strategy, settings, etc. The automatic proposal and selection of tools and machining parameters will be generated in accordance with the process of the expert's or experts' decision coded in the knowledge-based system. The user can access the contextual information and the selected rules and reasoning process, in order to justify the proposed solution. Confidence in the system and its proposals increases. Moreover, if any changes or new elements have to be implemented, all the structures and procedures already exist. All the maintenance and life of the knowledge-based systems are then available for the knowledge base or the software development.

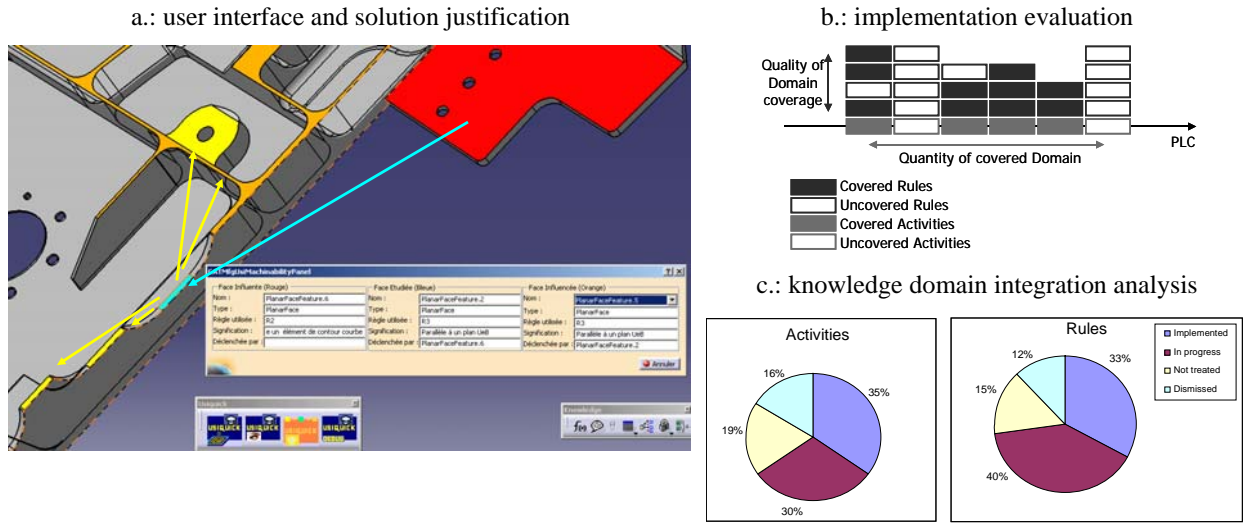

**Figure 4:** Knowledge based working environment and the implementation monitoring and traceability indicators

Due to the diversity of the engineering knowledge and the complexity of building KBE systems, it is difficult for the actors to evaluate if all the knowledge that should be automated has been taken into account, because a traditional development systems approach is based upon the realization of digital mock-ups.

But, by separating the activities of the capture phase from those of the formalisation phase, they could have at their disposal a first structured knowledge model and thus compare the two models.

This need of comparison introduces the need of knowledge traceability. This means that the capitalization process has to take into account the organizational aspect of the project in addition to the product and the process aspects.

To consider this new aspect an analysis of the developed algorithms has been done. The objective was to establish the correspondence between the algorithms and the design process activities in order to determine which activities have been effectively developed and, for each activity, the percentage of domain activities that has been automated (figure 4-c).

For this analysis the attribute "State" has been attached to each knowledge object to identify its state at a given time. The state can have one of the four following values: in progress, implemented, dismissed (ruled out: the implementation of the object is not envisaged), not treated.

## 7. *Conclusion*

Most KBE applications have been developed for solving large design problems in the aerospace and automotive industries where the main concern is the functionality to automate a complex design problem, rather than the reusability of engineering knowledge by the human expert.

However, to get such a result, disparate know-how and heterogeneous viewpoints have to be managed, integrated and stored in different forms that should be easily accessible, usable and maintainable. Ontology approaches can propose solutions that could help integrate knowledge in KBE environments.

The USIQUICK experience has shown that considering the two knowledge aspects separately, the content and the form, helps to decrease the complexity of knowledge-based engineering system development. The capitalization process we propose aims to structure knowledge engineering activities deployment.

It also aims to help the knowledge engineer capture all the knowledge he has in order to capitalize and to facilitate the communication between the different experts (or actors) and to have indicators regarding the project's lifecycle.